\documentclass[journal=jpclcd]{achemso}%

\usepackage{dcolumn}% Align table columns on decimal point
\usepackage{bm}% bold math
\usepackage[version=4]{mhchem}
\usepackage{siunitx}
\usepackage{subfigure}
\usepackage{amsfonts,amsmath,amssymb, subfigure}
\usepackage{color}
\usepackage{commath}%
\usepackage{esint}%
\usepackage[T1]{fontenc}
\usepackage{graphicx}
\usepackage[utf8]{inputenc}
\usepackage{microtype}
\usepackage{times}
\usepackage[textsize=footnotesize,obeyFinal]{todonotes}
\usepackage{xspace}
\usepackage[colorlinks=false]{hyperref}
\usepackage{hypernat}
\usepackage[version=4]{mhchem}%
\usepackage{textcomp}
\usepackage{diagbox}

\makeatletter
\renewcommand*{\acs@tocentry@print@aux}{%
  \begingroup
    \let\@startsection\acs@startsection@orig
    \acs@section*{\tocentryname}%
    \tocsize
    \sffamily
    \singlespacing
    \begin{center}
          \begin{minipage}{\acs@tocentry@height}
            \vbox to \acs@tocentry@width{\acs@tocentry@text}%
          \end{minipage}%
    \end{center}%
  \endgroup
}
\makeatother

\def\fig#1{Fig.~\ref{#1}}
\def\tab#1{Table.~\ref{#1}}

\def\ket#1{|#1\rangle}

\def\BK#1#2#3{\langle#1|#2|#3\rangle}

\newcommand{\mk}[1]{$#1$}
\newcommand*{\rep}{\ensuremath{\vec{r}\,^{\prime}}\xspace}
\newcommand*{\re}{\ensuremath{\vec{r}}\xspace}
\usepackage{color}
\definecolor{oldtxtcolor}{rgb}{0.00, 0.0, 0.5}
\definecolor{newtxtcolor}{rgb}{0.00, 0.3867, 0.00}
\definecolor{newtxtcolor}{rgb}{0.00, 0.0, 1}
\definecolor{oldtxtcolor}{rgb}{1.00, 0.0, 0.00}

\def\verX{12}
\def\verO{1}
\def\verN{2}
\def\verON{12}

\ifx\verX\verO
 \newcommand { \oldtxt }[1] {{\color{oldtxtcolor}{#1}}}
 \newcommand { \newtxt }[1] {}
\fi
\ifx\verX\verN
 \newcommand { \oldtxt }[1] {}
 \newcommand { \newtxt }[1] {{\color{newtxtcolor}{#1}}}
\fi
\ifx\verX\verON
 \newcommand { \oldtxt }[1] {{\color{oldtxtcolor}{#1}}}
 \newcommand { \newtxt }[1] {{\color{newtxtcolor}{#1}}}
\fi

\title{Identification of the decay pathway of  photoexcited nucleobases}
\author{Xiangxu~Mu*}
\affiliation{State Key Laboratory for Mesoscopic Physics and Collaborative Innovation Center of Quantum Matter, School of Physics, Peking University, Beijing 100871, China}
\author{Ming~Zhang*}
\affiliation{State Key Laboratory for Mesoscopic Physics and Collaborative Innovation Center of Quantum Matter, School of Physics, Peking University, Beijing 100871, China}
\author{Jiechao~Feng}
\affiliation{State Key Laboratory for Mesoscopic Physics and Collaborative Innovation Center of Quantum Matter, School of Physics, Peking University, Beijing 100871, China}
\author{Hanwei Yang}
\affiliation{State Key Laboratory for Mesoscopic Physics and Collaborative Innovation Center of Quantum Matter, School of Physics, Peking University, Beijing 100871, China}
\author{Nikita~Medvedev}
\affiliation{Institute of Physics Czech Academy of Science  Na Slovance 2, 182 21 Prague 8, Czech Republic}
\author{Xinyang~Liu}
\affiliation{State Key Laboratory for Mesoscopic Physics and Collaborative Innovation Center of Quantum Matter, School of Physics, Peking University, Beijing 100871, China}
\author{Leyi~Yang}
\affiliation{State Key Laboratory for Mesoscopic Physics and Collaborative Innovation Center of Quantum Matter, School of Physics, Peking University, Beijing 100871, China}
\author{Haitan~Xu}
\email{xuht@sustech.edu.cn}
\affiliation{Shenzhen Institute for Quantum Science and Engineering, Southern University of Science and Technology, Shenzhen 518055, China}
\alsoaffiliation{School of Physical Sciences, University of Science and Technology of China, Hefei 230026, China}
\author{Zheng~Li}
\email{zheng.li@pku.edu.cn}
\affiliation{State Key Laboratory for Mesoscopic Physics and Collaborative Innovation Center of Quantum Matter, School of Physics, Peking University, Beijing 100871, China}
\alsoaffiliation{{Collaborative Innovation Center of Extreme Optics, Shanxi University, Taiyuan, Shanxi 030006, China}}
\altaffiliation{Peking University Yangtze Delta Institute of Optoelectronics, Nantong, China}

\date{\today}
\begin{document}

\begin{tocentry}
\includegraphics[width=1.0\linewidth]{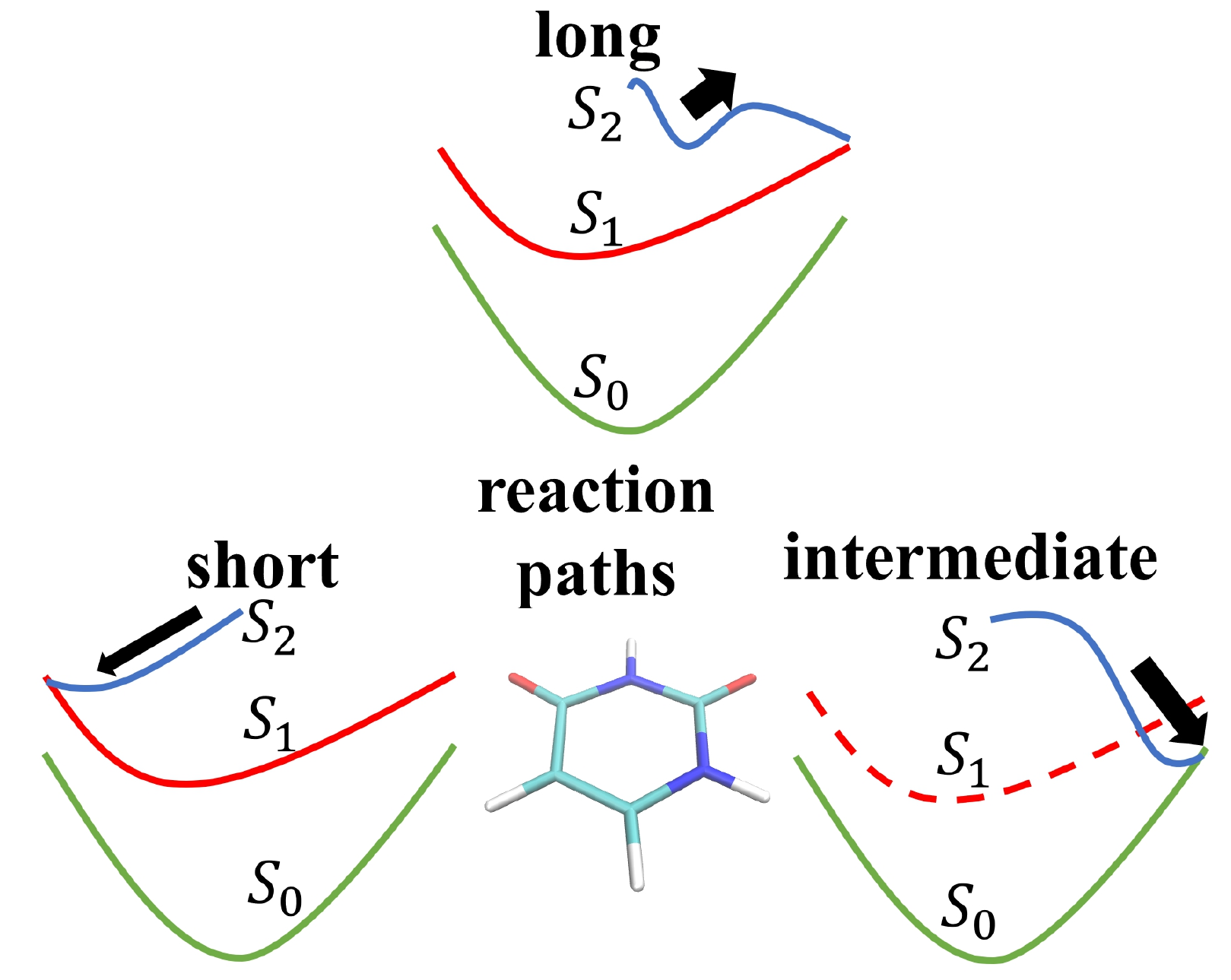}
\end{tocentry}

\begin{abstract}
The identification of the decay pathway of the nucleobase uracil after being photoexcited by ultraviolet (UV) light has been a long standing problem. Various theoretical models have been proposed but yet to be verified. Here we propose an experimental scheme to test the theoretical models by a combination of ultrafast X-ray spectroscopy,  X-ray diffraction and  electron diffraction methods. Incorporating the signatures of multiple probing methods, we demonstrate an approach to identify the pathway of the geometric and electronic relaxation of the photoexcited uracil molecule.
\end{abstract}
\maketitle

\begin{figure}[htb!]
    \centering
    \includegraphics[width=0.8\linewidth]{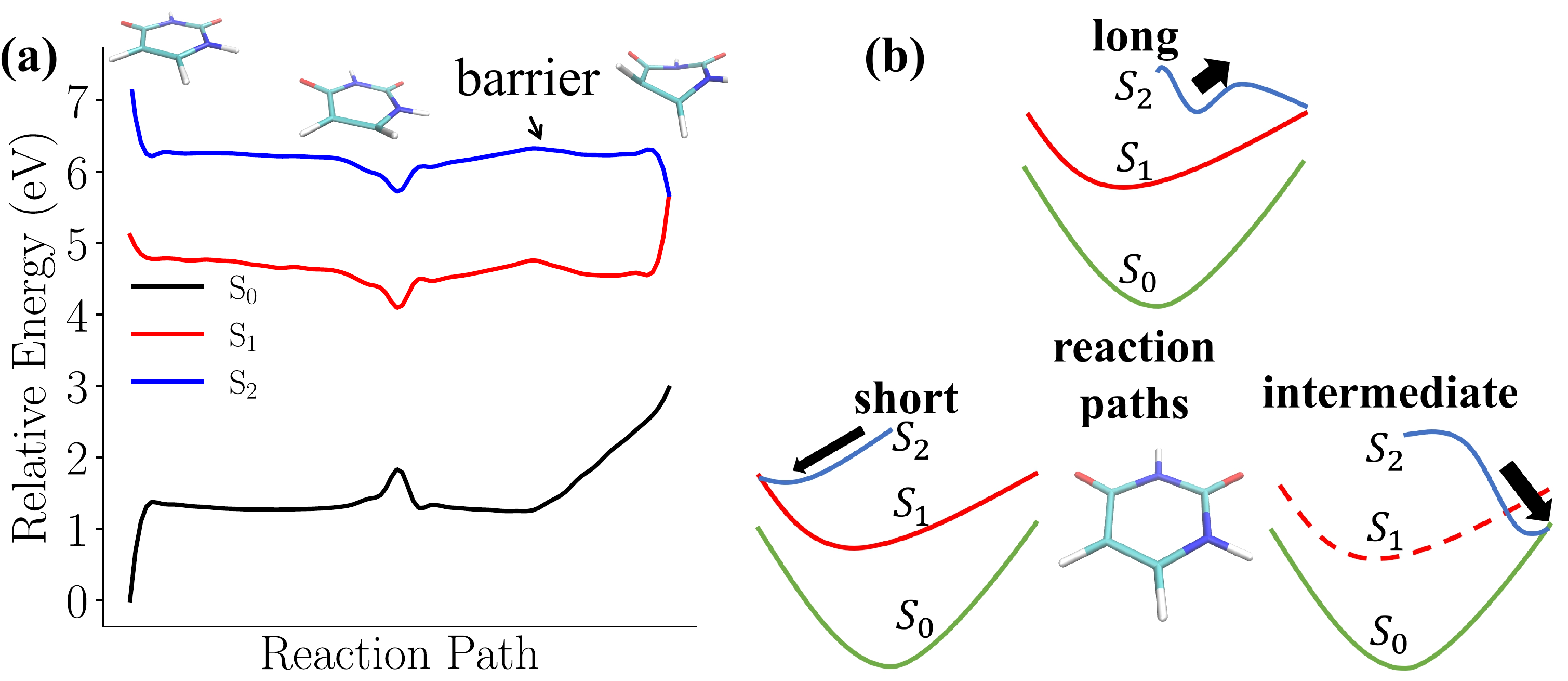}
    \caption{Predicted pathways for the electronic decay of photoexcited uracil. (a) Potential energy curves (PEC) for the reaction path of the long pathway, including the ME (S\mk{_0}), ME (S\mk{_2}), and MECI (S\mk{_2}/S\mk{_1}) geometries~\cite{hudock_ab_2007}, which are determined by the nudged elastic band (NEB) method~\cite{berne1998classical}, and the geometries are optimized on the SA5-CASSCF(8,6)//6-31g* level.
    There is a barrier of about 0.2 eV from ME (S\mk{_2}) to MECI (S\mk{_2}/S\mk{_1}).
    (b) Sketches of the three hypotheses of reaction paths and the equilibrium geometry of the uracil in the ground state.
    The long trajectory hypothesis assumes that the uracil relaxes into minimum energy geometry in the S\mk{_2} state within about 100 fs and then reaches the minimal energy conical intersection (MECI) between S\mk{_2} and S\mk{_1} states in several picoseconds.
    The short trajectory hypothesis assumes that the uracil arrives at S\mk{_1} in about 70 fs~\cite{lan2009Photoinduced}.
    The intermediate trajectory hypothesis assumes part of the uracil evolves to a CI point between S\mk{_2} and S\mk{_0} states within about 0.7~ps~\cite{nachtigallova2011Nonadiabatic}.
    }
    \label{fig:pes}
\end{figure}

%How will nature protect bio-molecules from photodamage?
Ultraviolet photons in the sunlight can excite biological  molecules, and the photoexcited molecules then experience different interatomic potential energies, which may induce unexpected reactions, such as dimethylation of RNA and DNA molecules, and seriously harm the biological functions of the molecules~\cite{smith1997Methylation, steenken1992Oneelectronreduction}.
In order to survive from the photodamage, living things seem to have chosen a special set of molecules as building blocks, which can decay rapidly at an ultrafast time scale via nonadiabatic pathways before harmful reactions take place~\cite{wolf_observation_2019, prokhorenko2016New}.
However, surprisingly, it was proposed that uracil, one of the nucleobases, could have a different property when being photoexcited to the singlet state S\mk{_2} by UV light.
Theoretical investigations showed that uracil may have a significantly longer electronic decay time up to picoseconds (ps) from the photoexcited state, because of a hypothetical barrier blocking the pathway to the conical intersection (CI) between the S$_2$/S$_1$ states (see \fig{fig:pes}(a))~\cite{hudock_ab_2007}. CI is a diabolical point in the potential energy surface caused by point-wise degeneracy of different electronic states and provides an ultrafast route of nonadiabatic electronic decay~\cite{Yarkony96:RMP68,Boggio02:JACS124, Graham04:ARPC55,Domcke12:ARPC63}.
The instability of RNA due to the long decay time of photoexcited uracil may result in gene mutations and evolution of life.
On the other hand, the proposal of ps long decay time of uracil is challenged by the follow-up studies~\cite{nachtigallova2011Nonadiabatic, lan2009Photoinduced,chakraborty2021Time, fingerhut2013Monitoring,hua2019Transient, richter_ultrafast_2014,brister2015Direct, nam2021Conical}, because the predicted potential barrier of $\sim$0.2 eV is very shallow, and due to the precision limit of quantum chemical calculations, different methods give contradictory predictions of electronic decay pathways.

The controversial predictions cover various time scales of electronic decay from the photoexcited S$_2$ state (see \fig{fig:pes}(b)). The long trajectory hypothesis~\cite{hudock_ab_2007} assumes that the relaxation is a two-step process. After being excited to S$_2$ from the ground state at the Franck-Condon (FC) region, the uracil first takes $\sim$ 100 fs to relax to a deformed geometry of minimal energy (ME) in the S$_2$ state, and then reaches the minimal energy conical intersection (MECI) between S$_2$ and S$_1$ states for the electronic decay, which could take picoseconds (ps) because of the potential energy barrier of about 0.2 eV, as shown in \fig{fig:pes}(a).
{The short trajectory hypothesis assumes that the uracil decays to S\mk{_1} in about 70 fs, and the nonadiabatic transition follows rapidly for an undistorted geometry}~\cite{lan2009Photoinduced}.
The intermediate trajectory hypothesis points to the third possibility of the decay pathway. {The uracil } can partially  circumvent the barrier and evolve to a CI point between S$_2$ and S$_0$ within $\sim$ 0.7 ps, which is energetically not favored but can result in direct transition  to the ground state S$_0$~\cite{nachtigallova2011Nonadiabatic}, and the intermediate state S$_1$ does not participate in this pathway.

Here we propose an approach to resolve the debate,  which can uniquely identify the electronic decay mechanism of the photoexcited uracil  by means of ultrafast X-ray spectroscopy and coherent diffraction imaging.
{We demonstrate that the combined ultrafast spectroscopic and diffraction signals can unambiguously distinguish  the decay  models from each other}.
The ultrafast electron diffraction (UED) is  capable of characterizing the evolving electronic correlation~\cite{yang2020Simultaneous} and molecular geometry~\cite{yang2018Imaging}, and can be used to monitor the electronic population transfer and transient structural dynamics~\cite{yangImagingCFConical2018, wolf2019photochemical}.
The ultrafast X-ray diffraction (UXD), though less sensitive to electron correlation, is free of pulse length limitation of UED due to space charge effect of electron bunch compression. For UXD with attosecond time resolution~\cite{duris2020Tunablea, duris2021Controllable}, it can resolve the transient geometric structure with  higher temporal precision.
The X-ray photoelectron spectroscopy (XPS) equipped with the ultrashort X-ray pulses from free electron lasers (FEL) provides the toolkit to map out the valence electron density variation in the chosen atomic sites of molecules in the excited state.
Incorporating  the mixed quantum-classical surface hopping molecular dynamics  (MD) method~\cite{tully_molecular_1990}, we simulate the trajectories that follow the long trajectory hypothesis, using the ab initio five-state-averaged complete active space self-consistent field method with 8 active electrons in 6 orbitals (SA5-CASSCF(8,6)) and 6-31g* basis set~\cite{hudock_ab_2007}, in order to show that a joint analysis based on UED, UXD and XPS data can test this hypothesis.
The surface hopping MD simulation of photoexcited uracil and the calculation of spectral and diffraction observables are carried out using the SHARC package~\cite{richter_sharc_2011, richter_ultrafast_2014}, and the quantum chemistry packages Molpro~\cite{MOLPRO2012} and Terachem~\cite{TERACHEM}, respectively (see details of the MD simulation in Supplementary Information [SI]).

\begin{figure}[htb!]
    \centering
    \includegraphics[width=0.75\linewidth]{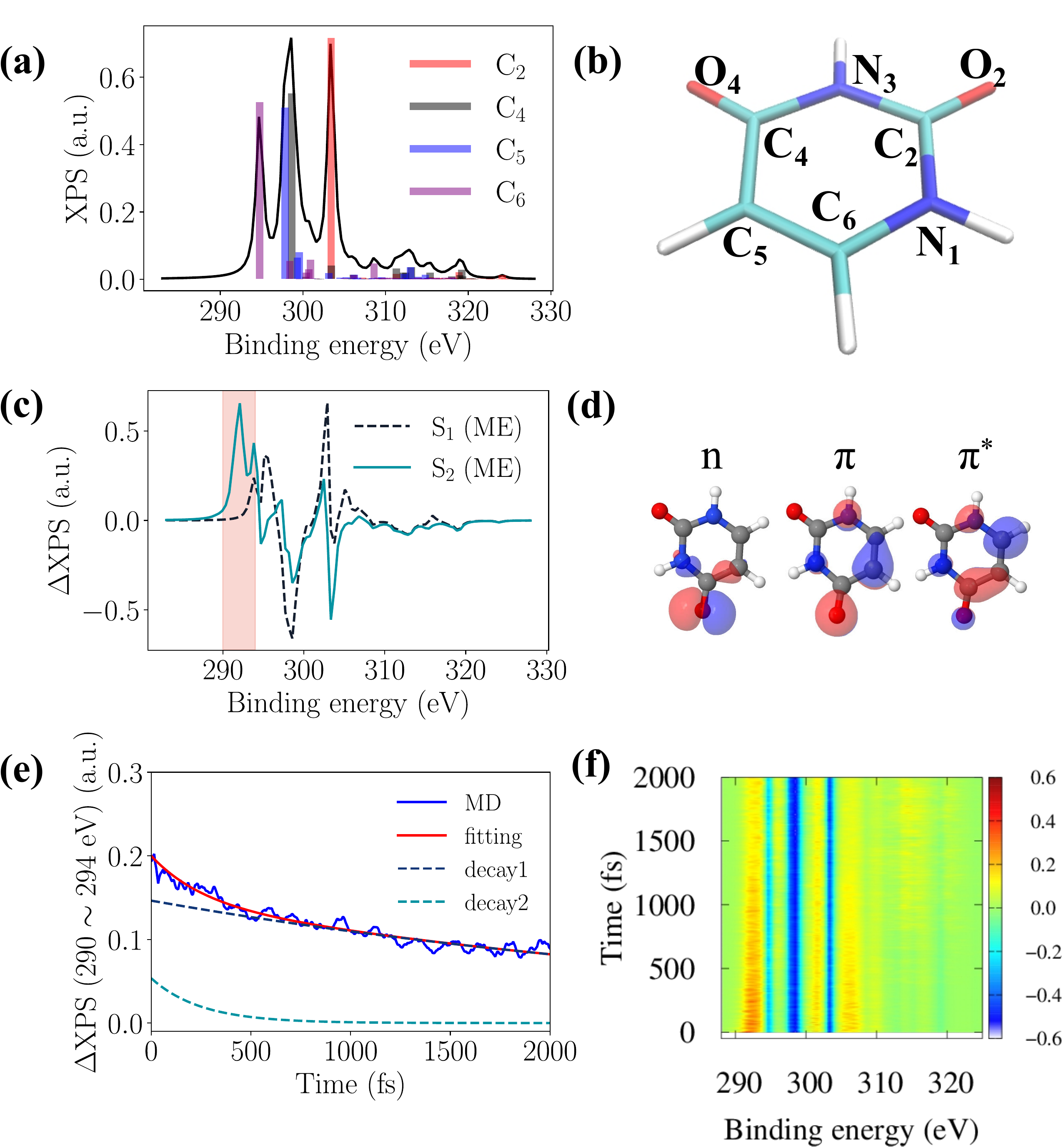}
    \caption{Calculated  XPS signals and their variations (\mk{\Delta}XPS) with respect to the electronic state transitions of uracil. 
    (a) XPS of ground state. The colored columns mark the intensity of transition to ionized final states, where the 1s\mk{^{-1}} hole is located at one of the four individual carbon atoms with different K-shell binding energies.  (b) Geometry with minimum energy. 
    %Black line is spectroscopy with 0.5 eV Lorentz line width. 
    (c) \mk{\Delta}XPS of minimum energy geometry of S\mk{_1} (dashed line) and S\mk{_2} (cyan line)  states relative to that of ground state S\mk{_0}.
    (d)  Molecular orbitals that are mostly relevant to the excited S\mk{_1} and S\mk{_2} states with \mk{n\pi^*} and \mk{\pi\pi^*} characters.
    (e) Temporal evolution of \mk{\Delta}XPS intensity integrated over the energy range from 290 to 294 eV with MD simulation (blue curve) and bi-exponential fitting (red curve). The two components of the bi-exponential fitting are also plotted, with decay time constants of 249~fs (black dashed curve) and 3469~fs (cyan dashed curve).
    (f) Energy resolved temporal evolution of \mk{\Delta}XPS intensity with MD simulation.
    }
    \label{fig:xps}
\end{figure}

Choosing the carbon K-edge for the X-ray probe, the shift of photoelectron energy of XPS in the molecule compared to that of carbon atom reflects the strength of electron screening of nuclear charge~\cite{siegbahn1967esca}, and maps out the local density of valence electrons at the specific atom, from which the 1s core electron is ionized, as shown in \fig{fig:xps}(a).
Because the excitation to the S\mk{_1} and S\mk{_2} states is accompanied by the flow of valence electrons from the non-bonding ($n$) and bonding $\pi$ orbitals to the antibonding $\pi^*$ orbital of uracil, the deficit and excess of valence electron density on the four individual carbon atoms gives the blue- and red-shift of carbon 1s binding energies in the XPS spectra, respectively (see \fig{fig:xps}(b)).
In the SI, we present the XPS spectra of uracil in the states involved in the MD simulation, including the singlet and triplet excited states.
{The Mulliken charge analysis of several representative geometries in the long trajectory hypothesis of electronic decay dynamics is shown in \tab{tab:xps_charge}. The most evident change is the increase of electronic density on \ce{C5} atom in the S\mk{_2} state, which leads to the red-shift of carbon 1s binding energy followed by the positive peak at 290 eV and the negative peak at 300 eV of \mk{\Delta}XPS of S\mk{_2}, as shown in \fig{fig:xps}(c).}
 {The change of \mk{\Delta}XPS intensity around 290 eV (shown as shaded area of \fig{fig:xps}(c)) can be uniquely linked to the evolution population of excited state S\mk{_2}, because the contribution of S\mk{_1} in this spectral range is negligible.}

\begin{table}[htb]
    \centering
    \begin{tabular}{|c|cccc|}
        \hline
        \diagbox{Structure}{Atom}    & 
        \ce{C2}   & \ce{C4}   & \ce{C5}   & \ce{C6} \\ \hline\hline
        FC (S\mk{_0})        & +1.05     & +0.81     & -0.32     & +0.17 \\ \hline
        FC (S\mk{_2})        & +1.05     & +0.75     & -0.53     & +0.13 \\ \hline
        ME (S$_{1}$)         & +1.08     & +0.69     & -0.33     & +0.05 \\ \hline
        ME (S$_{2}$)         & +1.04     & +0.77     & -0.46     & +0.12 \\ \hline
    \end{tabular}
    \caption{The {Mulliken} charge of {four individual} carbons atoms of Franck-Condon (FC) geometry in the S$_0$ and S$_2$ states and minimal energy (ME) geometries in the S$_1$ and S$_2$ states, the local charge deficit and excess upon geometric and electronic state variation leads to the blue and red shift of corresponding binding energies in XPS spectra, respectively.}
    \label{tab:xps_charge}
\end{table}

{However, apart from the transition of electronic states, the change of molecular geometry can also affect the spectral shift and intensity of \mk{\Delta}XPS, which mixes with the effect from the transition of electronic states and thus prohibits an unambiguous mapping of time-dependent XPS signals to the electronic population evolution (see simulated XPS of different states and geometries in the SI).}
To quantitatively extract the characteristic time constants of {the electronic decay out of the S$_2$ ($\pi\pi^*$) state and nuclear relaxation in the decay of photoexcited  uracil}, we apply bi-exponential fitting on \mk{\Delta}XPS$(t)$~\cite{richter_ultrafast_2014,mcfarland2014Ultrafast},
\begin{equation}
    \Delta\mathrm{XPS}(t) =A_\mathrm{XPS}[ N e^{-\frac{t}{\tau_1}} + (1-N) e^{-\frac{t}{\tau_2}}]
    \,,
\end{equation}
where \mk{A_\mathrm{XPS}} is the initial intensity of the \mk{\Delta}XPS$(t)$ signal, and $N$ quantifies the relative components of the geometric and electronic relaxation processes (see details of the fitting procedure in SI).
{However, the two time constants \mk{\tau_1\mathrm{(XPS)} = 249} fs and \mk{\tau_2\mathrm{(XPS)} = 3468 } fs can not be unambiguously assigned to the characteristic time scale of geometric relaxation \mk{T_\alpha} and electronic decay \mk{T_\beta}, as the fitting model makes no assumptions about the physics of the temporal trajectory, but only  quantifies the time scales of the reactions~\cite{gao2013Mapping}.
%As the signatures of the two relaxation processes mix in the same spectral region, the long and short time constants are not necessarily mapped to the electronic and geometric relaxation time, respectively.
%
In order to resolve this difficulty, we propose a multi-signal analysis using spectroscopic (XPS) and diffraction (UED and UXD) signals to investigate these two processes and validate the two time constants, %using the signatures in the ultrafast diffraction, 
which can be uniquely assigned to the different types of relaxation processes.}

\begin{figure}[htb]
    \centering
    \includegraphics[width=1.0\textwidth]{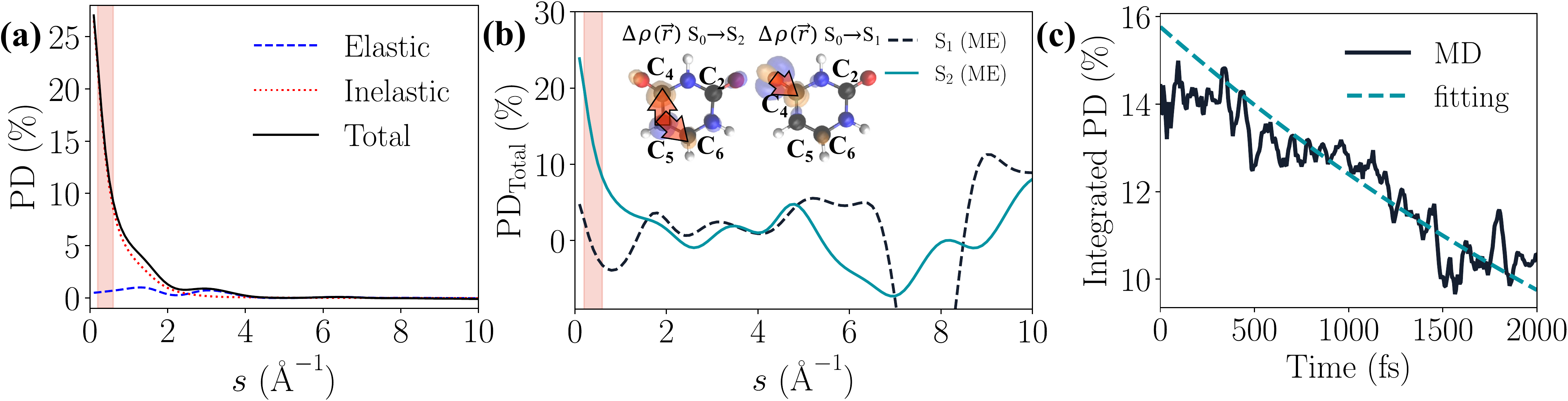}
    \caption{Simulated ultrafast electronic diffraction (UED) signal for uracil. 
    (a) Percentage difference (PD) of the UED signals of uracil at the Franck-Condon geometry in the S\mk{_2} state. Compared to elastic signal (blue dashed line), inelastic signal (red dotted line) contributes predominantly to the total signal (black line) in the small-angle region.  
    (b)  PDs of the total signals for S\mk{_1} and S\mk{_2} states of the minimum energy structures.
    Shaded areas in (a) and (b) correspond to the small-angle scattering region \mk{0.2<s<0.6}\AA\mk{^{-1}}. 
    Inset of (b) sketches the variation of electron density \mk{\rho(\re)} upon transition from S$_0$ to  \mk{n\pi^*}  (S\mk{_1}) and \mk{\pi\pi^*} (S\mk{_2}) states.
    The orange and blue colors correspond to the positive and negative isosurface \mk{\Delta \rho(\re)} of $0.01$\AA$^{-3}$.
    (c) Temporal evolution of small-angle PD\mk{_\mathrm{Total}}($t$) of the total signal.
    The decay time constant  is \mk{\tau\mathrm{(UED)}=4166} fs from exponential fitting of PD\mk{_\mathrm{Total}}($t$) .}
    \label{fig:uedinela}
\end{figure}

{Ultrafast electron diffraction (UED) provides a tool for retrieving transient molecular structural and electronic dynamics simultaneously, and exhibits high sensitivity for measuring electronic correlations from small angle scattering signals~\cite{yang2020Simultaneous}. }
The intensities of the elastic and inelastic scattering signals are
%%%%%%%%%%%%%%%%%%%
\begin{equation}
    I_\mathrm{elastic}(\vec{s}) = \frac{1}{s^4} |
    \sum_\alpha N_\alpha e^{i \vec{s}\cdot\vec{R}_\alpha} - f(\vec{s})
    |^2
    \,,
    \label{eq:uedela}
\end{equation}
\begin{equation}
    I_\mathrm{inelastic}(\vec{s}) = \frac{1}{s^4} [
    n + P(\vec{s}) - \left | f(\vec{s})\right |^2
    ]
    \,,
    \label{eq:uedinela}
\end{equation}
%
%%%%%%%%%%%%%%%%%%%
where \mk{\vec{s} = \vec{k}_\mathrm{in} - \vec{k}_\mathrm{out}} is the momentum transfer of electrons, \mk{N_\alpha} and \mk{\vec{R}_\alpha} are the nuclear charge and position of the \mk{\alpha}-th atom, \mk{n} is the number of electrons in the molecule, \mk{f(\vec{s})} and \mk{P(\vec{s})} are the Fourier transforms of one-electron density \mk{\rho(\re)} and two-electron density \mk{\rho^{(2)}(\re,\rep)}, 
%%%%%%%%%%%%%%%%%%%
\begin{equation}
    f(\vec{s}) = \int e^{i\vec{s}\cdot\re} \rho(\re) d\re
    \,,
\end{equation}
\begin{equation}
    P(\vec{s}) = \int e^{i \vec{s} \cdot (\re - \rep)}
    \rho^{(2)}(\re, \rep) d \re d \rep
    \,.
\end{equation}
%%%%%%%%%%%%%%%%%%%
%
{The inelastic scattering intensity \mk{I_\mathrm{inelastic}(\vec{s})} dominates at small scattering angles as shown in the shaded area of \fig{fig:uedinela}(a). Because the inelastic electron scattering is dependent on the Fourier transform of the two-electron density $ \rho^{(2)}(\re, \rep) $, it measures the  changes in the electron  correlation due to the transitions of electronic states}.
In contrast, the elastic scattering signal \mk{I_\mathrm{elastic}(\vec{s})} dominates at larger scattering angles, and encodes the transient structural information {characterized by the atomic charge pair distribution functions (CPDF)}~\cite{wolf2019photochemical,yang2021Structure}, which is given by
\begin{eqnarray}
\mathrm{CPDF}(R,t)=R \int_0 ^{s_\mathrm{max}} s^5 I(s,t) e^{-\alpha s^2} \sin (s R) ds
\,,
\end{eqnarray}
where $I(s,t)$ is the isotropic average of total UED signal including both elastic and inelastic components. The damping term \mk{e^{-\alpha s^2}} with \mk{\alpha = 0.06} is introduced to avoid edge effects during the transform~\cite{yang2021Structure}.
We define the percentage difference (PD) of the UED signal as
\begin{eqnarray}
\mathrm{PD} = \frac{I_\mathrm{UED} - I_{0,\mathrm{UED}}}{I_{0,\mathrm{UED}}} \times 100\%,
\end{eqnarray}
where $I_{0,\mathrm{UED}}$ is the UED signal in the equilibrium geometry of the ground state and $I_\mathrm{UED}$ is that of excited states.
The S\mk{_1} state is mainly from the excitation of the localized non-bonding $n$ orbital to delocalized {C-O antibonding} \mk{\pi^*} orbital relative to the ground state S$_0$ (see \fig{fig:uedinela}(b)), which is accompanied by the enlarged two-electron distance and thus the reduction of electronic Coulomb repulsion and the electron correlation. Such process must result in the enhancement of inelastic scattering signals at small scattering angles, thus S$_1$ has weaker electron correlation than that of S$_0$.
The excitation to S\mk{_2} state accompanied by the transition between two delocalized orbitals \mk{\pi} and \mk{\pi^*} leads to longer range electron flow around the molecular ring. As shown in \fig{fig:uedinela}(b), the electrons relocate from \ce{C5} atom to two nearest neighbor atoms \ce{C4} and \ce{C6}, which forms a more delocalized electron density distribution and results in larger PD of inelastic signal than S\mk{_1} state in the small $s$ region.

Due to the evident PDs of the  inelastic scattering signals for S\mk{_2} (\mk{\sim}20\%) and S\mk{_1} ($<$10\%) at small angles (0.2\mk{<s<}0.6\AA$^{-1}$) (see \fig{fig:uedinela}(b)) as well as for other relevant states (see SI), the inelastic signal can serve as a sensitive probe for the transition of electronic states.
We show the PD of the inelastic signal calculated from MD trajectories in \fig{fig:uedinela}(c), and fit  PD(\mk{t}) with an exponential function as
$
 \mathrm{PD}(t) = A e^{-\frac{t}{\tau}}
 \,.
$
The time constant extracted from the fitting is \mk{\tau\mathrm{(UED)} = 4166} fs, which qualitatively matches the magnitude of electronic decay time constant by time-resolved XPS analysis \mk{\tau_2}(XPS), and correctly reflects the corresponding parameter of the long trajectory model.
%
%The error of decay time constant from UED signals may come from the minor contribution from the S$_1$ state in the overlapped spectral region.
%
However, the fast geometric relaxation~\cite{hudock_ab_2007} could pose a challenge to UED, because its time resolution is partially limited by the space charge effect of the electron pulses.
On the other hand, the sub-100~fs structural dynamics can be well resolved by UXD using ultrashort X-ray pulses from X-ray free electron lasers (XFEL), which can reach an attosecond time resolution~\cite{duris2020Tunablea,duris2021Controllable}.  

\begin{figure}[htb!]
    \centering
    \includegraphics[width=0.6\linewidth]{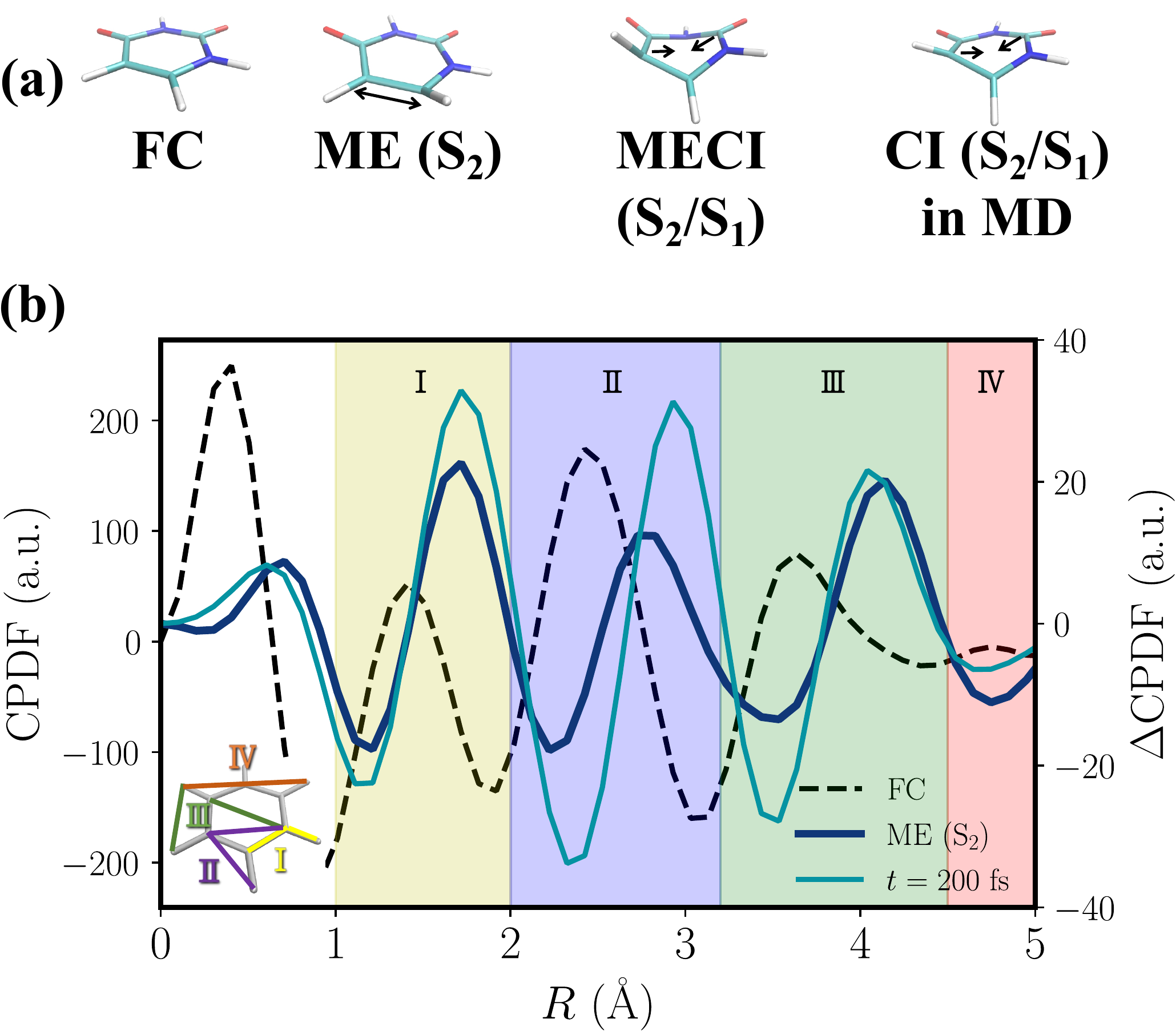}
    \caption{Charge pair distribution function (CPDF) of static structures and MD trajectories.
    (a) Geometries of FC, ME (S\mk{_2}), MECI (S\mk{_2}/S\mk{_1}) and a representative CI point.
    The black arrows depict the major pairwise deformation of the geometries of the equilibrium FC point.
    (b) CPDF signal of FC (black dashed curve), variation of CPDF signal of ME (S\mk{_2}) (blue solid curve) and that of the molecular geometry at the representative time  \mk{t = 200} fs (cyan solid curve).
    The yellow, purple, green and red areas correspond to the 1st shell (\mk{1.0<R<2.0} \AA), 2nd shell (\mk{2.0<R<3.2} \AA), 3rd shell (\mk{3.2<R<4.5} \AA) and 4th shell (\mk{4.5<R<5.0} \AA). 
    They are marked by \uppercase\expandafter{\romannumeral1} (yellow), \uppercase\expandafter{\romannumeral2} (purple), \uppercase\expandafter{\romannumeral3} (green) and \uppercase\expandafter{\romannumeral4} (red), corresponding to those in the inset of (b).}
    \label{fig:cpdf}
\end{figure}
To reveal the molecular dynamical information in the simulated UXD data, we apply the spectral and autocorrelation analysis to the charge pair distribution functions (CPDF).
As shown in \fig{fig:cpdf}(b), the black line is the CPDF of uracil at equilibrium FC point.
The peak at \mk{R<1} \AA\xspace is contributed by inelastic diffraction, which reflects the electron-electron correlation.
The negative CPDF at \mk{R \simeq 1} \AA\xspace comes from the electron-nucleus pairs.
Longer range interactions dominate the other peaks, in which the structural information of the atomic positions in the molecule is encoded~\cite{yang2021Structure}. 
The  peak at \mk{R \simeq 1.4} \AA\xspace comes from the elastic scattering of the nearest neighbour (NN) atom pairs, dubbed the 1st shell.
The third peak (at \mk{R \simeq 2.5} \AA\xspace) comes from the next nearest neighbour (NNN) atomic pairs, dubbed the 2nd shell.
%%%%%%
The fourth peak at \mk{R \simeq 3.8} \AA\xspace corresponds to the second atomic coordination shell (the distances between atoms are two atomic sites; 3rd shell).
The fifth peak comes from the third coordination shell (the distances between atoms are three atomic sites; 4th shell).
%%%%%%
The CPDF of charge pairs of various shells are shown in  \fig{fig:cpdf}(b), marked by \uppercase\expandafter{\romannumeral1}(yellow), \uppercase\expandafter{\romannumeral2}(purple), \uppercase\expandafter{\romannumeral3}(green) and \uppercase\expandafter{\romannumeral4}(red), respectively.
The first three geometries in \fig{fig:cpdf}(a) are the same representative configurations as those in \fig{fig:pes}(a) along the reaction path in the S$_2$ state.
A representative CI structure in the MD trajectories is shown in \fig{fig:cpdf}(a).
The CPDF of other shells are shown in SI.
According to \fig{fig:cpdf}(b), as uracil moves toward ME (S\mk{_2}), the charge pair density in the 1st shell increases. 
These variations of shells manifest themselves clearly in the elastic scattering signal. As shown in the \mk{\Delta}CPDF signal at ME (S\mk{_2}) in \fig{fig:cpdf}(b), compared with CPDF signal at FC, the peak intensity of the 1st shell of CPDF moves toward  larger \mk{R}.
We also show the \mk{\Delta}CPDF signal at 200~fs from MD trajectories, which is a representative time point near  ME (S\mk{_2}). 
It exhibits similar positive peak at \mk{\sim 2} \AA\xspace and negative peak at \mk{\sim 1} \AA, and indicates the bond length of \ce{C5}-\ce{C6} atoms for ME (S\mk{_2}) that is the dominant driving reaction coordinate for geometrical relaxation of photoexcited uracil in the long trajectory model~\cite{hudock_ab_2007}.
These stretching modes of the \ce{C5} and \ce{C6} atoms are the focus of our quantitative analysis of the structural dynamics.
To capture the characteristic signatures of the structural evolution, we apply the continuous wavelet transform (CWT) analysis to the autocorrelation function of CPDF($t$) from the simulated time-resolved ultrafast X-ray diffraction (UXD) signal, which gives the frequency spectra of  nuclear motions with largest amplitudes at each time points.
The autocorrelation function \mk{A(t)} is given by
\begin{equation}
 A(t) = \int_{R_a}^{R_b} \mathrm{CPDF}(R,t)\times \mathrm{CPDF}(R,0)dR
 \,,
\end{equation}
calculated from a swarm of MD trajectories, where \mk{R_a}, \mk{R_b} are the boundaries of 1st atomic shell, and is detrended by a baseline \mk{A_\mathrm{f}(t)} fitted to the exponential model (see details in SI), giving
\begin{equation}
\label{eq:autocorr}
    A_\mathrm{detrend}(t) = A(t) - A_\mathrm{f}(t)
    \,.
\end{equation}
%
%%%%%%%%%%%%%%%%%%%%%%%%%%
\begin{figure}[htb!]
    \centering
    \includegraphics[width=0.9\linewidth]{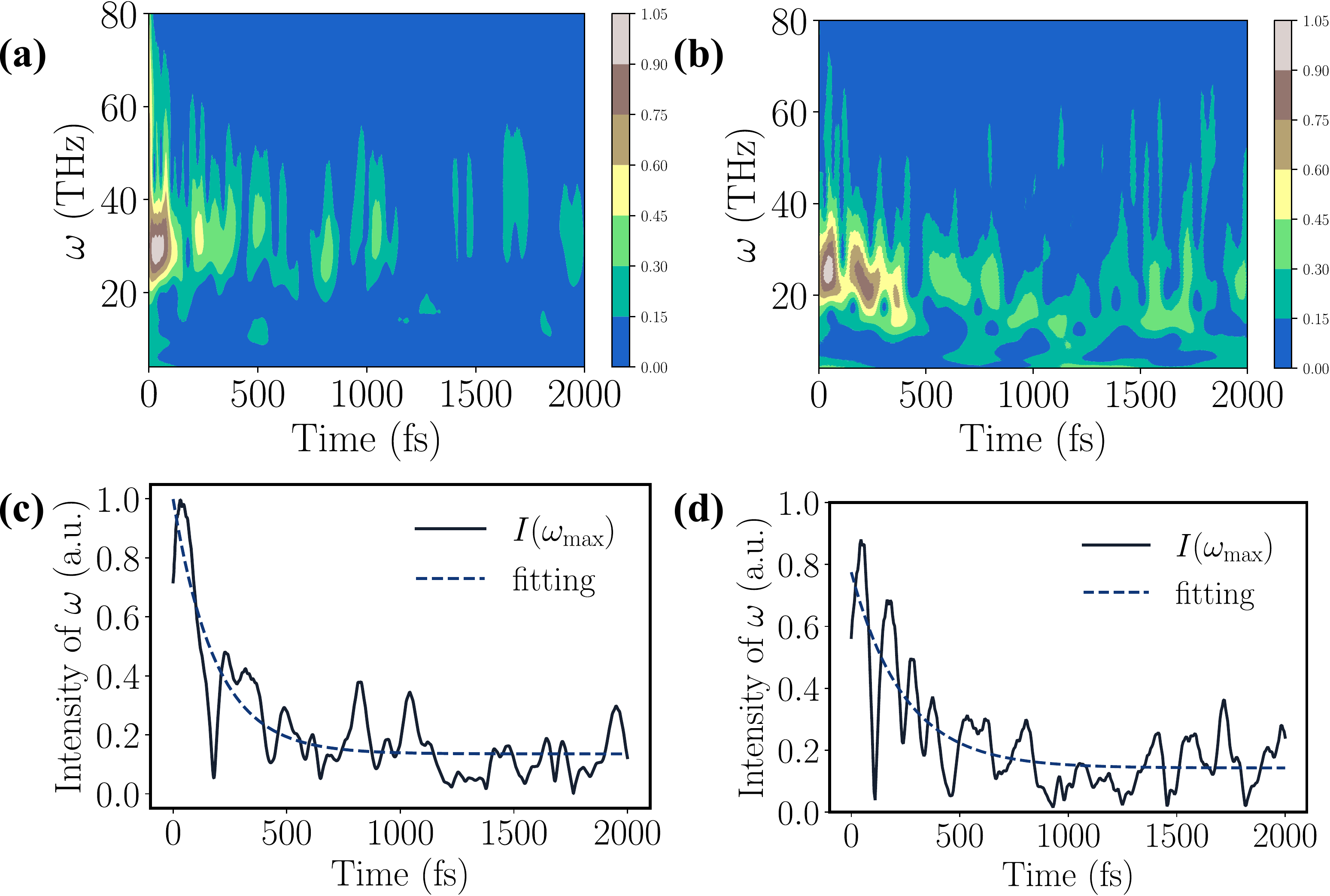}
    \caption {Dynamics of the geometrical relaxation of the photoexcied uracil. (a) Wavelet transform of autocorrelated bond length evolution of the \ce{C5}-\ce{C6} atoms from MD trajectories. (b) Wavelet transform of the detrended autocorrelation function of CPDF of ultrafast x-ray diffraction signal \mk{A_{\mathrm{detrend}}(t)} in Eq.~\ref{eq:autocorr}. (c) Temporal evolution of the vibrational amplitude of  the \ce{C5}-\ce{C6} bond at \mk{\omega_\mathrm{max} = 29.2} THz, with time constant of 185 fs by exponential fitting. \mk{\omega_\mathrm{max}} is the frequency where the wavelet transform is maximal when \mk{t = 0}. (d) Temporal evolution of \mk{A_{\mathrm{detrend}}} at \mk{\omega_\mathrm{max} = 29.2} THz, with time constant of 243 fs by exponential fitting. }
    \label{fig:cwt}
\end{figure}
%%%%%%%%%%%%%%%%%%%%%%%%%%
In the long trajectory model, the driving mode corresponds to the \ce{C5}-\ce{C6} stretching, which is active in the first 500 fs~\cite{hudock_ab_2007}.
As shown in the wavelet transformed spectra (\fig{fig:cwt}(a)) and the exponential fitting (\fig{fig:cwt}(c)) of the evolution of the \ce{C5}-\ce{C6} bond length from MD trajectories, the vibrational spectrum peaks around 30 THz at the beginning in the S$_2$ state for to a period of 33 fs, and this geometric stretching mode dominates around 185 fs.
The frequencies of the major modes match that in \fig{fig:cwt}(b), which is obtained from wavelet transform of autocorrelation function of the 1st shell. Time-dependent frequency spectra of nuclear motion in other shells are shown in the SI.
The time constant of the major modes (\fig{fig:cwt}(d)), $\sim$243 fs, is also obtained from the exponential fitting of wavelet transform of autocorrelation function of the 1st shell CPDF.
The analysis shows that the geometric relaxation time scale of $<$0.5 ps in the long trajectory model can be qualitatively obtained via the time-dependent frequency analysis of ultrafast diffraction with sufficient time resolution.

We have shown that the characteristic time scales of geometric relaxation and electronic decay in the long trajectory model can be faithfully retrieved by incorporating time-resolved XPS, UED and UXD analysis.
In the long trajectory model, the relaxation mechanism of photoexcited uracil is comprised of two  processes. One is the fast geometric relaxation with the characteristic time \mk{T_\alpha < 0.5} ps, and the other is a slower transition of the electronic state  to S$_1$ with a characteristic time \mk{T_\beta\sim 3-4} ps.
It is important to note that none of the three methodologies alone can determine the characteristic time constants of the two competing relaxation processes of uracil involving nuclear and electronic degrees of freedom. 
%Each of the methods provides data that are complementary to that from other methods, and the time constants can only be finally validated from their mutual consistency.

With the same method, one can test the intermediate and short trajectory hypotheses by experimental measurements.
These two hypotheses have one major process that is different from the two-step model in the long trajectory hypothesis. The molecule tends to directly find the decaying geometry in the S$_2$ state, without first evolving to the minimum energy geometry as in the long trajectory hypothesis.
In the short trajectory model (\fig{fig:pes}(b)), the characteristic time of the electronic decay from S\mk{_2} to S\mk{_1} is less than 100 fs and the molecule keeps the plane geometry, and the conical intersection is very close to the Franck-Condon region in the S$_2$ state~\cite{lan2009Photoinduced}.
In this case, the XPS spectral components of S$_2$ must decay within 100 fs, which can be observed both in the XPS and the inelastic signal of UED.
The CI point in the short trajectory model has a stretched \ce{C5}-\ce{C6} bond without the ring-folding characteristic, and the peaks of CPDF are expected to move toward larger  $R$ in sub-100 fs and a similar peak of \ce{C5}-\ce{C6} stretching oscillation should appear in the time-frequency analysis of the UXD autocorrelation function.
If the molecule decays along the intermediate pathway as shown in \fig{fig:pes}(b), photoexcited uracil will follow the S\mk{_2} state potential energy surface until reaching an ethylenic CI point between S\mk{_2} and S\mk{_0}~\cite{nachtigallova2011Nonadiabatic}, which  can then decay to the S\mk{_0} ground state in $\sim 700$ fs. 
In this process, it is expected that the geoemtric relaxation possesses the same time constant as the electronic decay, i.e. $\sim$ 700 fs, and can be mapped out  by XPS and the inelastic part of UED. The UXD offers complementary evidence for the structural dynamics in finding the S$_2$/S$_0$ intersection, which should be driven by a small set of reaction coordinates~\cite{gao2013Mapping}. Such driving coordinates can be revealed by the time-frequency analysis of UXD.

While the predicted phenomena have not yet been fully examined experimentally, they are within reach of the capabilities of free electron lasers and ultrafast electron diffraction facilities.
Our study demonstrates the synergy of the X-ray photoelectron spectroscopy, and the electron and X-ray diffraction  with ultrafast time resolution.
The approach can also serve as a general methodological toolkit for investigating valence electron and structural dynamics in ultrafast photochemistry.

\section{METHODS}
\subsection{X-ray photoelectron spectra}
Under dipole approximation, the ionization rate from the initial state $\ket{\psi_I}$ with energy $E_I$ is
\begin{equation}
    P(E)\propto \sum_{f,\eta}\left|\BK{\psi_f\phi_{\eta}}{\hat{\vec{\mu}}\cdot\vec{\mathcal{E}}}{\psi_I}\right|^2 \delta(\hbar\omega+E_I-E_f-E)
    \,,
\end{equation}
where the  final state includes the molecular cationic state $\ket{\psi_f}$ with energy $E_f$ and the ejected electron state $\ket{\phi_\eta}$ with energy $E$, $\hbar\omega$ is the X-ray photon energy, $ and \hat{\vec{\mu}}$ and $\vec{\mathcal{E}}$ are the electronic dipole operator and the electric field of X-ray. In the second-quantized form,
\begin{equation}
    \BK{\psi_f\phi_\eta}{\hat{\vec{\mu}}\cdot\vec{\mathcal{E}}}{\psi_I}=\sum_{ij}\BK{\psi_f\phi_\eta}{\hat{a}^\dagger_i\hat{a}_j}{\psi_I} \vec{\mu}_{ij}\cdot\vec{\mathcal{E}}=\sum_{p}\BK{\psi_f}{\hat{a}_p}{\psi_I} \vec{\mu}_{\eta p}\cdot\vec{\mathcal{E}}
    \,,
\end{equation}
where $i,j$ are orbital indexes including all bounded and continuum states, and orbital index $p$ is restricted to be the  carbon 1s orbital. $\vec{\mu}_{ij}$ and $\vec{\mu}_{\eta p}$ are matrix elements of transition dipole moment between corresponding initial and final orbitals. Replace $\vec{\mu}_{\eta p}$ by their average value $\vec{\mu}_{\eta p}\cdot\vec{\mathcal{E}}\approx f(E)$ which is only dependent of the energy of the ejected electron $E$ , then
\begin{equation}
    P(E)\propto \sum_{f}\left|\sum_p\BK{\psi_f}{\hat{a}_p}{\psi_I}\right|^2 \rho(E)f(E) \delta(\hbar\omega+E_I-E_f-E)
    \,.
\end{equation}
The sum over all ejected electron states is proportional to the factor $\rho(E)$, which is the density of states. For photoionization in low energy region far from resonance, the product $\rho(E)f(E)$ can be treated as a constant~\cite{hudock_ab_2007}, so
\begin{equation}
    P(E)\propto \sum_{f}\left|\sum_p\BK{\psi_f}{\hat{a}_p}{\psi_I}\right|^2 \delta(\hbar\omega+E_I-E_f-E)
    \,.
\end{equation}
The calculation involves three initial states S\mk{_0}, S\mk{_1} and S\mk{_2}, and different cationic final states with 1s\mk{^{-1}} hole located at different carbon atoms are considered separately, as shown in \fig{fig:xps}(a). For each carbon 1s\mk{^{-1}} hole, 50 cationic final states with lowest energy are involved in the calculation.

\begin{suppinfo}
% Detail of molecular dynamics, \textcolor{red}{table of contents of SI to be completed}
The SI contains the initial conditions for molecular dynamics, population dynamics, X-ray photoelectron spectrum (XPS) for various electronic states and molecular geometries, fitting result of time resolved XPS, ultrafast electron diffraction analysis with comparison between S\mk{_2} state at FC point and S\mk{_2} state at ME point, and time-frequency analysis of various atomic shells.
\end{suppinfo}

\begin{acknowledgement}
This work was supported by the National Natural Science Foundation of China (No. 12174009, 11974031).
We thank Todd J. Martinez, Fang Liu, Basile Curchod and Ludger Inhester for helpful discussions.
NM gratefully acknowledges financial support from the Czech Ministry of Education, Youth and Sports (grants No. LTT17015, LM2018114, and EF16\_013/0001552).
\end{acknowledgement}

\section*{Author Contributions}
M.X.X. and M.Z. contributed equally to this work.

\bibliography{references}

\end{document}

%
% Numbering all lines
% \setpagewiselinenumbers
% \modulolinenumbers[1]
% \linenumbers

\title{Supplementary Information for \\ Identification of the decay pathway of  photoexcited nucleobases}
\author{Xiangxu~Mu*}
\affiliation{State Key Laboratory for Mesoscopic Physics and Collaborative Innovation Center of Quantum Matter, School of Physics, Peking University, Beijing 100871, China}
\author{Ming~Zhang*}
\email{These authors contribute equally to this work.}
\affiliation{State Key Laboratory for Mesoscopic Physics and Collaborative Innovation Center of Quantum Matter, School of Physics, Peking University, Beijing 100871, China}
%
% \author{Basile~Curchod}
% \affiliation{Chemistry Department, Durham University, UK}
%
\author{Jiechao~Feng}
\affiliation{State Key Laboratory for Mesoscopic Physics and Collaborative Innovation Center of Quantum Matter, School of Physics, Peking University, Beijing 100871, China}
% \author{Fang~Liu}
% \affiliation{Chemistry Department, Emory University, USA}
\author{Hanwei Yang}
\affiliation{State Key Laboratory for Mesoscopic Physics and Collaborative Innovation Center of Quantum Matter, School of Physics, Peking University, Beijing 100871, China}
% \author{Ludger~Inhester}
% \affiliation{Center for Free-Electron Laser Science, DESY, Notkestrasse 85, 22607 Hamburg, Germany}
\author{Nikita~Medvedev}
\affiliation{Institute of Physics Czech Academy of Science  Na Slovance 2, 182 21 Prague 8, Czech Republic}
\author{Xinyang Liu}
\affiliation{State Key Laboratory for Mesoscopic Physics and Collaborative Innovation Center of Quantum Matter, School of Physics, Peking University, Beijing 100871, China}
\author{Leyi~Yang}
\affiliation{State Key Laboratory for Mesoscopic Physics and Collaborative Innovation Center of Quantum Matter, School of Physics, Peking University, Beijing 100871, China}
\author{Haitan~Xu}
\email{xuht@sustech.edu.cn}
\affiliation{Shenzhen Institute for Quantum Science and Engineering, Southern University of Science and Technology, Shenzhen 518055, China}
\affiliation{School of Physical Sciences, University of Science and Technology of China, Hefei 230026, China}
% \author{Todd~J.~Martinez}
% \affiliation{Department of Chemistry and the PULSE Institute, Stanford University, 333 Campus Drive, Stanford, California 94305, USA}
% \email{todd.martinez@stanford.edu}
%
\author{Zheng~Li}
\email{zheng.li@pku.edu.cn}
\affiliation{State Key Laboratory for Mesoscopic Physics and Collaborative Innovation Center of Quantum Matter, School of Physics, Peking University, Beijing 100871, China}
\affiliation{{Collaborative Innovation Center of Extreme Optics, Shanxi University, Taiyuan, Shanxi 030006, China}}
\affiliation{Peking University Yangtze Delta Institute of Optoelectronics, Nantong, China}
\date{\today}

\maketitle

\section{Non-adiabatic Tully Surface Hopping (TSH) Molecular Dynamics}
\subsection{Preparation of initial conditions for TSH simulations}
\begin{figure}[htb!]
    \centering
    \includegraphics[width=0.6\linewidth]{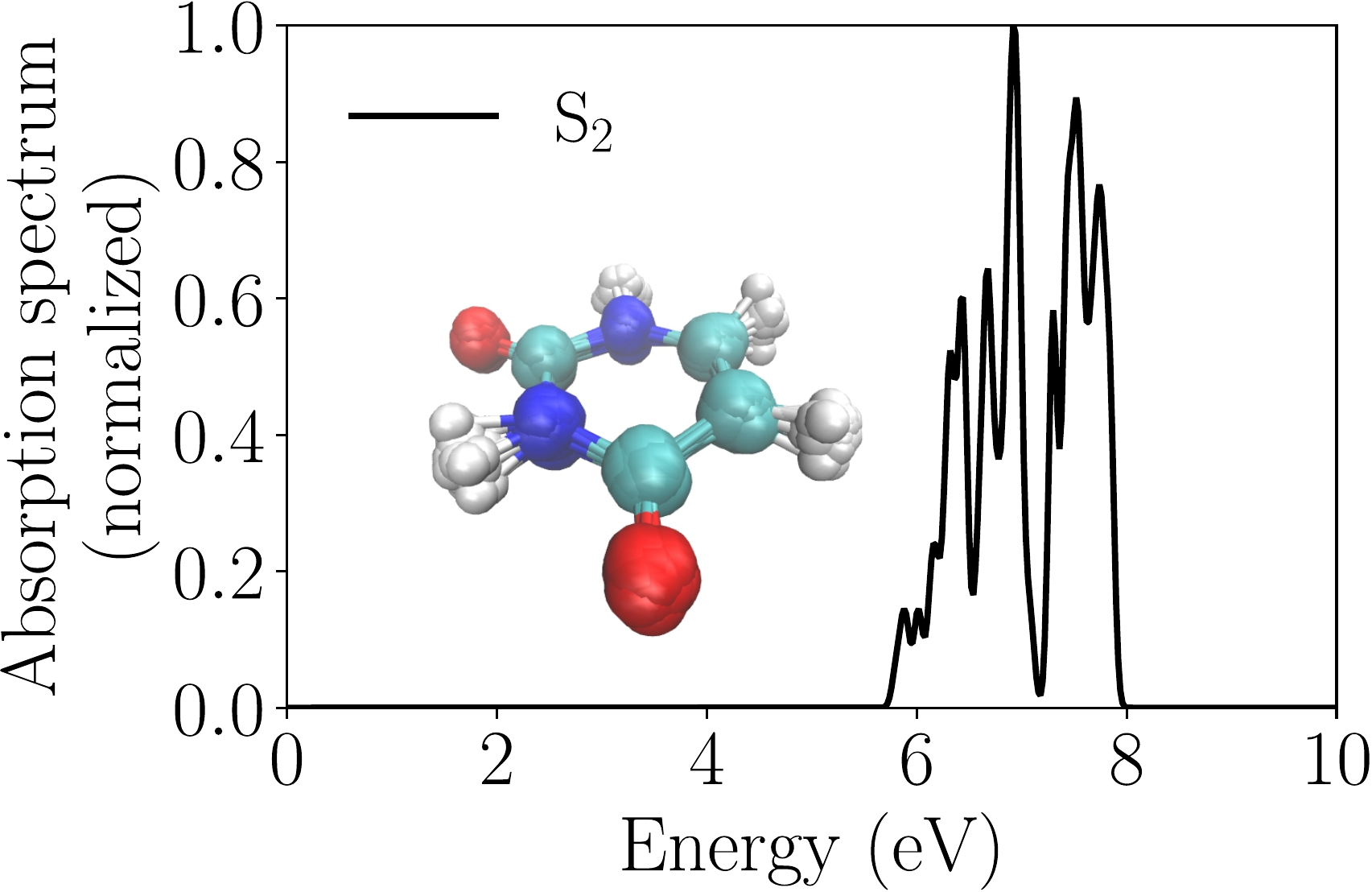}
    \caption{UV absorption spectrum. The UV absorption spectrum was simulated with initial conditions sampled from a ground state Wigner distribution.}
    \label{sfig:init}
\end{figure}

An ultraviolet (UV) absorption spectrum (see \fig{sfig:init}) is generated from geometries sampled from a Wigner distribution corresponding to the optimized ground state structure~\cite{hudock_ab_2007}. 
Single point energy calculations are performed for the singlet states at the SA5-CASSCF(8,6)/6-31G\mk{^*} level for all  the initial geometries and their \mk{\mathrm{S}_0 \rightarrow \mathrm{S}_2} excitation energies homogeneously broadened using Gaussian functions with a full width at half maximum (FWHM) of 0.1 eV. 
Positions and momenta for all these different initial conditions used to generate the electronic absorption spectra are used to initiate the TSH dynamics from the \mk{\mathrm{S}_2} state.
%%%%%%%%%%%%%%%%%%%%%
%%%%%%%%%%%%%%%%%%%%%
\subsection{Population Dynamics}
\begin{figure}[htb!]
    \centering
\includegraphics[width=0.6\linewidth]{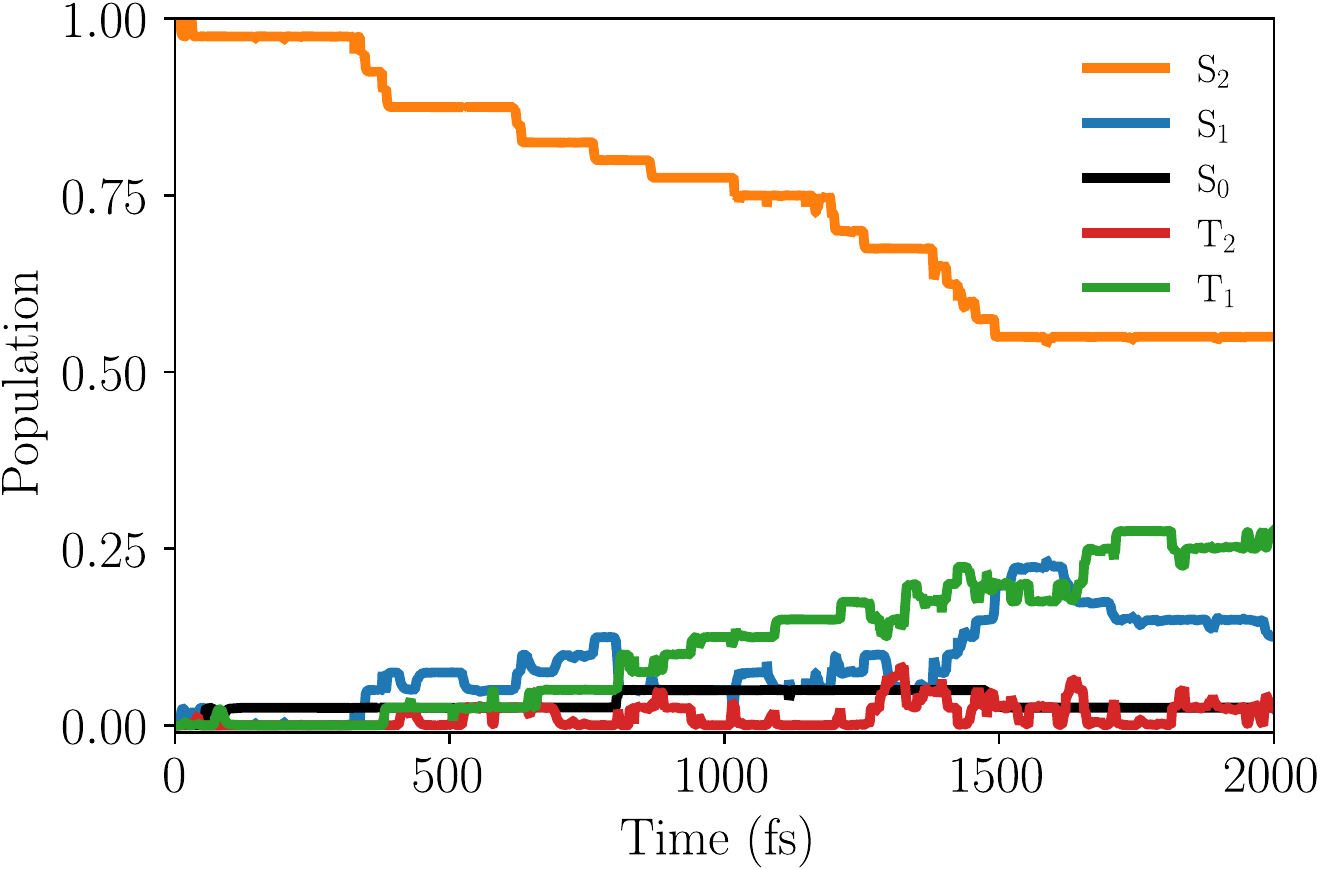}
    \caption{Population dynamics for uracil in the first 2 ps of the TSH simulations.}
    \label{sfig:popu}
\end{figure}

The population dynamics of the nuclear wave packet of all electronic states are shown in  \fig{sfig:popu}. Uracil exhibits an incubation period~\cite{wolf2019photochemical} in the long trajectory model, where essentially no population is transferred to the other states for \mk{\sim 400} fs. The population in the \mk{\mathrm{S}_2} state decays slowly, with 50\% population transferred to the \mk{\mathrm{S}_1} state after more than 2 ps.

\section{XPS of uracil in the ground and excited states}
In this section, we discuss the capability and limitation of XPS to map out (\mk{\tau_1}(XPS),\mk{\tau_2}(XPS)) with the electronic state relaxation time and the structural relaxation time (\mk{T_ \alpha}, \mk{T_\beta}).
%%%%%%%%%%%%%%%%%
\begin{figure}
    \centering
    \includegraphics[width=0.9\linewidth]{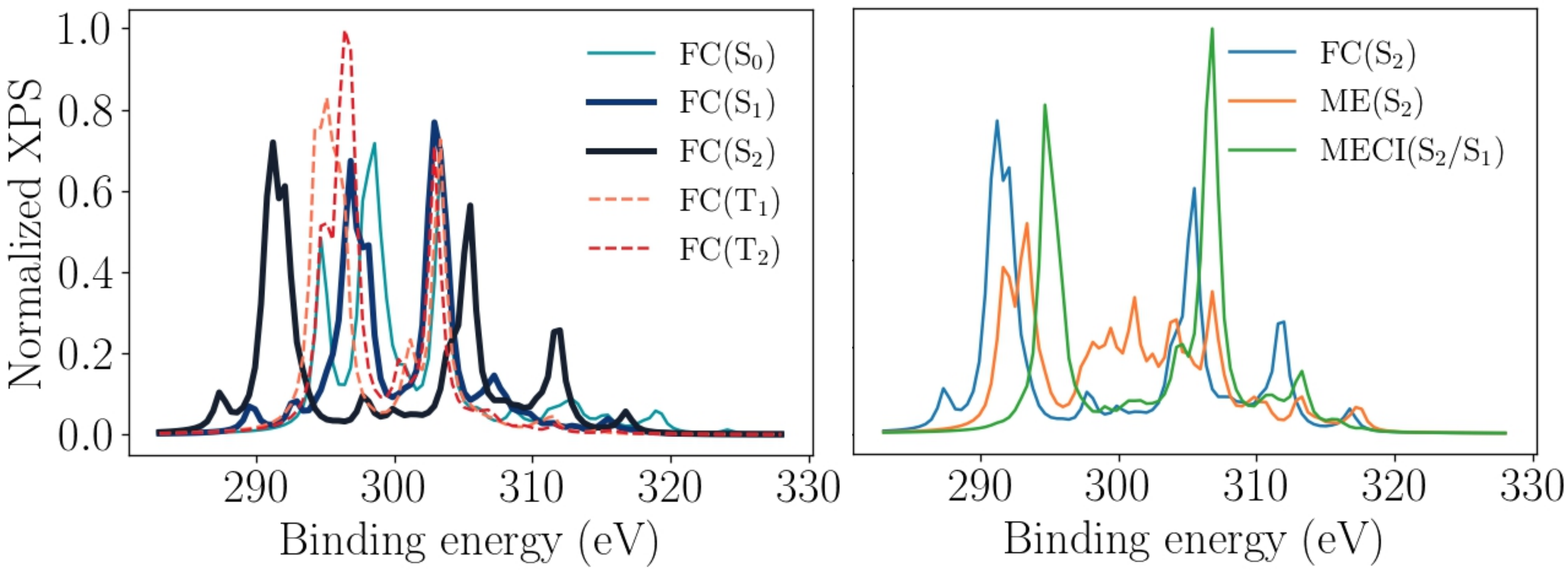}
    \caption{Characterization of the XPS spectra.}
    \label{sfig:xps}
\end{figure}
%%%%%%%%%%%%%%%%%
As shown in \fig{sfig:xps}(a), at FC, the only large peak at \mk{\sim 291} eV belongs to the S\mk{_2} state. This is an evidence that XPS encodes the information of electronic states. Nevertheless, when uracil evolves to MECI from FC, even for the same state, the peak at \mk{\sim 290} eV is reduced, as shown in \fig{sfig:xps}(b).

\subsection{Fitting Model for XPS }
\begin{figure}[htb]
    \centering
    \includegraphics[width=0.6\linewidth]{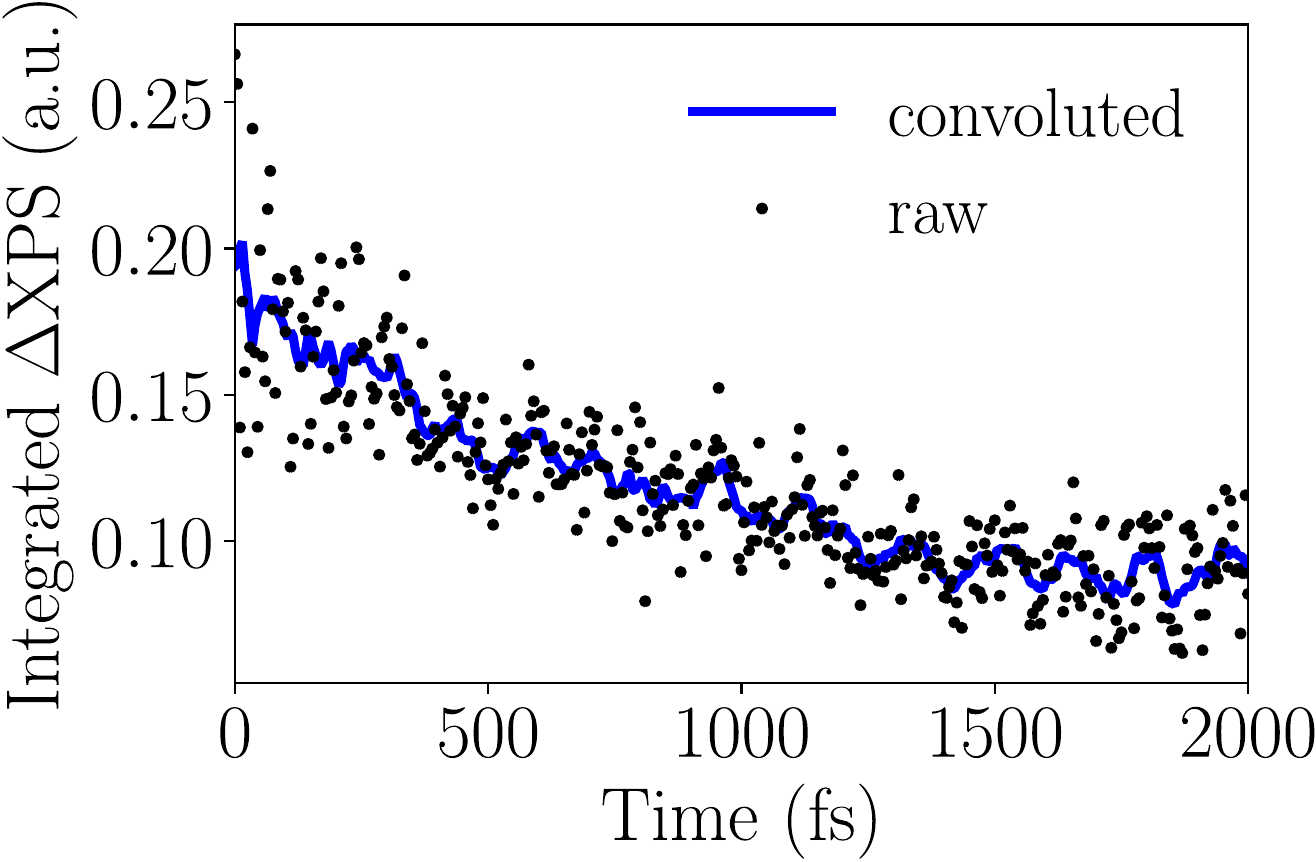}
    \caption{Intensity of integrated $\Delta$XPS as a function of time. 
    The black points are the raw data of the integrated $\Delta$XPS from 290 to 294 eV.
    The blue curve is the convoluted result.}
    \label{sfig:xpst}
\end{figure}

%The MD simulation has strong oscillation which may come from the vibrational mode.
The raw data of simulated time-resolved XPS from MD trajectories are  convoluted for the fitting and extraction of time constants, which are shown in \fig{sfig:xpst}. 
The raw data are convoluted with the package \textit{tsmoothie}~\cite{pelagatti2015time}, which integrates the raw data and weights  generated using a linear window function. 
The fitting parameters of XPS are shown in \tab{stab:fitxps}.
\begin{table}[!ht]
    \centering
    \begin{tabular}{ccccc}
    \hline
        Parameters & \mk{A_\mathrm{XPS}} & $N$ & \mk{\tau_1}(XPS) (fs) & \mk{\tau_2}(XPS) (fs) \\ \hline\hline
        Results & 0.2 & 0.732 & 3468.76 & 249.13 \\ \hline
    \end{tabular}
    \caption{Bi-exponential fitting result of the integrated \mk{\Delta}XPS.}
    \label{stab:fitxps}
\end{table}

\section{Ultrafast Electron Diffraction }
\begin{figure}[htb!]
    \centering
    \includegraphics[height=1.0\textwidth]{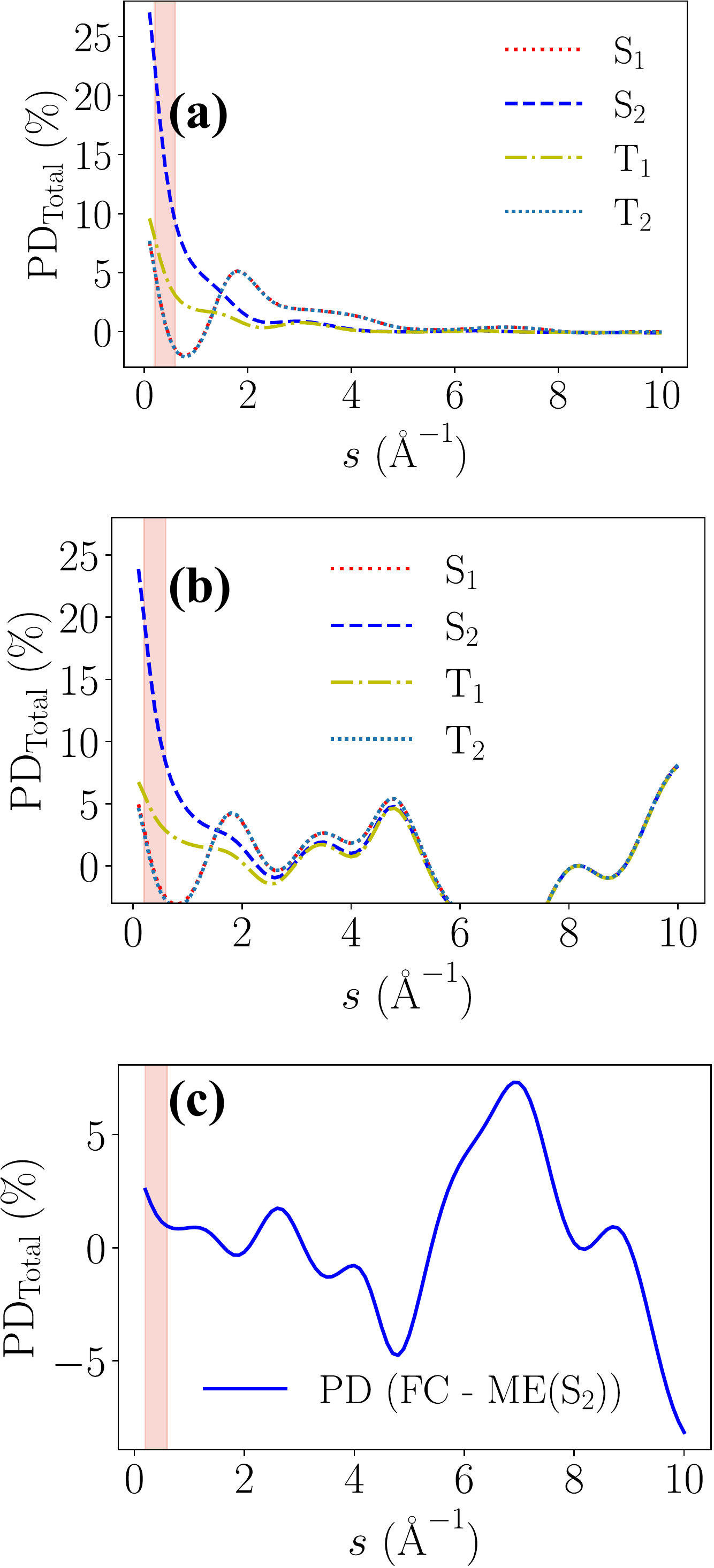}
    \caption{The PD signal of various states with different geometries.
    (a)  PD signal of the three singlet states and the two triplet states at the FC point.
    (b) PD signal at the ME (S\mk{_2}) point.
    (c)  Difference between the PD signal of S\mk{_2} state at the FC point and that at  the ME~(S\mk{_2}) point.
    }
    \label{sfig:pd}
\end{figure}

As shown in \fig{sfig:pd}, the PD signal in the small scattering angle region of S\mk{_2} state is insensitive to the deformation of uracil geometry. There is only \mk{\sim}2\% PD difference at small scattering angles between two representative structures in the same state (see \fig{sfig:pd}(c)), and thus the inelastic electron scattering can be used to map out the electron correlation variation uniquely.

\section{Wavelet Transform and Time-Frequency Analysis}
\begin{figure}[htb]
    \centering
    \includegraphics[width=0.8\linewidth]{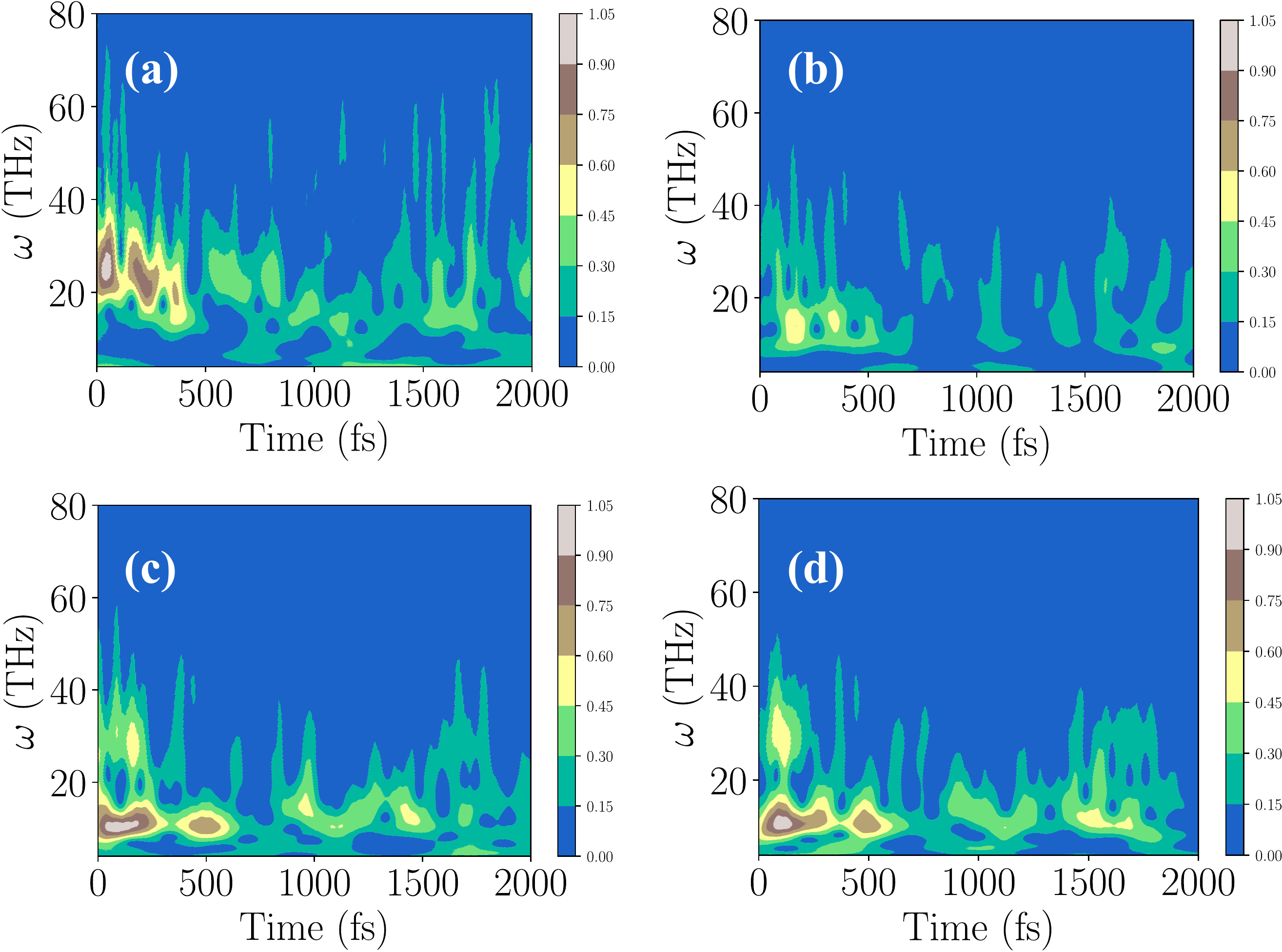}
    \caption{Wavelet transform of the autocorrelated CPDF of four shells for time-frequency analysis of MD trajectories.}
    \label{sfig:cwtall}
\end{figure}

The time-frequency analysis of MD trajectories using wavelet transform of the autocorrelated CPDF of four shells are given in \fig{sfig:cwtall}, from which we can conclude that the \ce{C5}-\ce{C6} bond vibration is the dominant mode. This mode has a frequency of $\sim$ 30 THz and is active for the first 500 fs.
%
%%%%%%%%%%%%%%%%%%%%%%%%%%
%%%%%%%%%%%%%%%%%%%%%%%%%%

\bibliography{references}